\UseRawInputEncoding
\documentclass[aps,prb,citeautoscript,floatfix,twocolumn,superscriptaddress,bibnotes]{revtex4-1}
\usepackage{physics}
\usepackage{graphicx}
\usepackage{xcolor}
\usepackage[pdftex,bookmarks=true,bookmarksopen,bookmarksnumbered,
                colorlinks,
                linkcolor=blue,
                citecolor=blue]{hyperref}

\usepackage{float}
\usepackage[geometry]{ifsym}
\newcommand{\NN}{$^{14}$N}
\newcommand{\CC}{$^{13}$C}
\newcommand{\nv}{NV$^-\:$}

\begin{document} 

\title{Observing hyperfine interactions of \texorpdfstring{NV$^-$}{NV-} centers in diamond in an advanced quantum teaching lab}
\author{Yang Yang}
\affiliation{
 School of Electrical Engineering and Telecommunications,
 The University of New South Wales, Sydney, NSW 2052, Australia
}
\affiliation{
Quantum Photonics Laboratory and Centre for Quantum Computation and Communication Technology, School of Engineering, RMIT University, Melbourne, Victoria 3000, Australia
}

\author{Hyma H. Vallabhapurapu}
\affiliation{
 School of Electrical Engineering and Telecommunications,
 The University of New South Wales, Sydney, NSW 2052, Australia
}

\author{Vikas K. Sewani}
\affiliation{
 School of Electrical Engineering and Telecommunications,
 The University of New South Wales, Sydney, NSW 2052, Australia
}

\author{Maya Isarov}
\affiliation{
 School of Electrical Engineering and Telecommunications,
 The University of New South Wales, Sydney, NSW 2052, Australia
}

\author{Hannes R. Firgau}
\affiliation{
 School of Electrical Engineering and Telecommunications,
 The University of New South Wales, Sydney, NSW 2052, Australia
}

\author{\\Chris Adambukulam}
\affiliation{
 School of Electrical Engineering and Telecommunications,
 The University of New South Wales, Sydney, NSW 2052, Australia
}

\author{Brett C. Johnson}
\affiliation{
Quantum Photonics Laboratory and Centre for Quantum Computation and Communication Technology, School of Engineering, RMIT University, Melbourne, Victoria 3000, Australia
}

\author{Jarryd J. Pla}
\affiliation{
 School of Electrical Engineering and Telecommunications,
 The University of New South Wales, Sydney, NSW 2052, Australia
}

\author{Arne Laucht}
\affiliation{
 School of Electrical Engineering and Telecommunications,
 The University of New South Wales, Sydney, NSW 2052, Australia
}

\begin{abstract}
The negatively charged nitrogen-vacancy (NV$^-$) center in diamond is a model quantum system for university teaching labs due to its room-temperature compatibility and cost-effective operation. 
Based on the low-cost experimental setup that we have developed and described for the coherent control of the electronic spin (Sewani et al.~[\onlinecite{Sewani2020}]), we introduce and explain here a number of more advanced experiments that probe the electron-nuclear interaction between the \nv electronic and the \NN~and \CC~nuclear spins. Optically-detected magnetic resonance (ODMR), Rabi oscillations, Ramsey fringe experiments, and Hahn echo sequences are implemented to demonstrate how the nuclear spins interact with the electron spins. Most experiments only require 15 minutes of measurement time and can, therefore, be completed within one teaching lab. 

\end{abstract}

\maketitle

\section{Introduction}

The recent progress in the field of quantum technologies -- quantum computing, quantum communications and quantum sensors -- has led to considerable commercial interest from entities like the semiconductor industry, software companies and consulting firms. Quantum technologies are predicted to constitute a $\sim$\$10 billion market within the next decade~\cite{Omdia2019}, and the high demand for quantum-specialists, as reflected by the high number of current job advertisements~\cite{QCReport2020}, corroborates this prediction. Universities need to cater to this need and provide their graduates with a thorough understanding of quantum mechanics and solid-state physics, including experimental experience in how to perform quantum measurements and interpret results. However, these skills are difficult to convey in a university teaching lab, due to the sensitivity of quantum systems to environmental disturbances, and the high cost of specialized measurement equipment. 

As discussed amply in literature, the nitrogen vacancy (NV$^-$) centre in diamond possesses an electronic spin that can be initialized and readout optically. It displays microsecond coherence times at room temperature in ambient conditions~\cite{Doherty2013}. This makes it a model system for high-school or university teaching labs to perform magnetometry~\cite{Zhang2018,Qutools,Misonou2020} and coherent spin control experiments~\cite{Bucher2019,Misonou2020,Sewani2020,Ciqtek,SpinFlex}. 
Based on the same low-cost experimental setup that we have developed in Ref.~\onlinecite{Sewani2020}, in combination with an off-the-shelf, commercially available, high-quality chemical vapour deposition (CVD) diamond sample from \emph{Element Six}, 
we expand the portfolio of experiments to include the characterization of electron-nuclear spin coupling dynamics.   

This paper is organized as follows: In the proceeding section (Section~\ref{sec_LO}) we will give an overview of the learning outcomes that can be conveyed with the experiments in this paper. Section~\ref{sec_NV} will then introduce the \nv Hamiltonian with a focus on the terms describing the nuclear spin of the \NN~nucleus and its interaction with the electronic spin, and Section~\ref{sec_RabiTheory} will introduce the Rabi formula that describes the time evolution of a spin during spin resonance. In Section~\ref{sec_Equip} we will detail the changes made to the equipment with regards to Ref.~\onlinecite{Sewani2020}, and in Section~\ref{sec_Exp} we will present the different experiments with a special emphasis on how the coupling of the electron spin to the \NN~nucleus manifests in the optically-detected magnetic resonance (ODMR) spectrum (Subsection~\ref{sec_Spec14N}), Rabi oscillations (Subsection~\ref{sec_Rabi}), and Ramsey fringes (Subsection~\ref{sec_Ram}). Finally, we show how the coupling to the bath of \CC~nuclei manifests in the Hahn echo decay (Subsection~\ref{sec_Hahn}).

\section{Learning outcomes}\label{sec_LO}
The experiments described below are a set of advanced experiments that build upon the basic ones described in Ref.~\onlinecite{Sewani2020}. The basic experiments taught the students fundamental concepts such as the \nv center structure, the spin initialization and readout procedure, the features of the ODMR spectrum, development of pulse sequences, the rotating frame, magnetic resonance, two-axes control, the Bloch sphere, and dynamical decoupling. 
We will assume knowledge of these concepts for this manuscript, and build upon them for the students to:
\begin{itemize}
    \item Develop a detailed understanding of the system's Hamiltonian and the ability to map the observed experimental features to its properties.
    \item Comprehend the effect of experimental non-idealities like power-broadening and spectral detuning.
    \item Observe the interaction between the \nv electronic spin and the \NN~nuclear spin, mediated by the hyperfine interaction.
    \item Experience how the electron-nuclear interaction manifests in both the frequency domain and the time domain, and how those are linked by the Fourier transform.
\end{itemize}

The concepts demonstrated with these experiments will provide the students with an advanced understanding of quantum experiments and a detailed knowledge of electron-nuclear spin interactions. They are applicable to state-of-the-art research on \nv centers in diamond~\cite{Childress2013,Abobeih2018,Abobeih2019}, other types of color centers in diamond~\cite{Bhaskar2020,Wan2020} and silicon carbide~\cite{Bourassa2020}, donor spins in silicon~\cite{Freer2017,Asaad2020,Madzik2020}, and even gate-defined quantum dots~\cite{Hensen2020}.

\section{The \texorpdfstring{NV$^-$}{NV-} center in diamond}\label{sec_NV}
The \nv ~center in diamond consists of a substitutional nitrogen atom adjacent to a vacant lattice site, and exists along four different crystallographic orientations ($[111]$, $[1\bar{11}]$, $[\bar{1}1\bar{1}]$ and $[\bar{11}1]$)~\cite{Doherty2013,Zhang2018,Sewani2020}. These four orientations are in principle equivalent, but lead to different alignments of the \nv center axes with respect to an externally applied static or oscillating magnetic field (see Fig.~\ref{figOrientation}). The natural abundance of \NN~is 99.6\% and its nuclear spin number is $I=1$, which means that the electron spin in almost all \nv centers is coupled to a nuclear spin. We therefore need to extend the ground state Hamiltonian of the \nv center to account for the coupled \NN~nuclear spin. The static Hamiltonian ${\cal H}_0$ contains three main terms:
\begin{equation}
\label{eq_H0}
    {\cal H}_0 ={\cal H_{\rm S}}+{\cal H_{\rm I}}+{\cal H_{\rm SI}},\\
\end{equation}
where ${\cal H_{\rm S}}$ is the electron spin Hamiltonian, ${\cal H_{\rm I}}$ is the nuclear spin Hamiltonian, and ${\cal H_{\rm SI}}$ describes the interaction of the electron spin with the nuclear spin.

The electron spin Hamiltonian
\begin{equation}
\label{eq_HS}
    {\cal H_{\rm S}} =D\,S_{\rm z}^2+\gamma_{\rm e}\,\mathbf{B_0}\cdot\mathbf{S}
\end{equation}
corresponds to the well-known Hamiltonian from Ref.~\onlinecite{Sewani2020}, where $D\approx2.87$~GHz is the zero-field splitting of the ground state $|{g}\rangle$, and \textbf{S} is the vector of electron spin matrices $[S_{\rm x}, S_{\rm y}, S_{\rm z}]$ for an $S=1$ system. The electron gyromagnetic ratio, which defines the magnitude of the Zeeman splitting, is given by $\gamma_{\rm e}=28$~GHz/T, and $\mathbf{B_0}$ is the external magnetic field vector. 
As in Ref.~\onlinecite{Sewani2020} we write the Hamiltonians in units of frequency, as this is more relevant to experiment. 

The nuclear spin Hamiltonian
\begin{equation}
\label{eq_HI}
    {\cal H_{\rm I}} =P\,I_{\rm z}^2-\gamma_{\rm n}\,\mathbf{B_0}\cdot\mathbf{I}
\end{equation}
is the corresponding term for the \NN~nuclear spin. $P\approx-5.01$~MHz is the nuclear quadrupole moment of \NN, \textbf{I} is the vector of nuclear spin matrices $[I_{\rm x}, I_{\rm y}, I_{\rm z}]$ for $I=1$, and $\gamma_{\rm n}=3.077$~MHz/T is the nuclear gyromagnetic ratio of \NN~\cite{PhysRevB.79.075203,phdthesis_Yossi,phdthesis_claudia}. Since $\gamma_{\rm n}$ is roughly a factor $10,000$ smaller than $\gamma_{\rm e}$, the interaction between the nuclear spin and the external $\mathbf{B_0}$ field is not observed in our experiments as it would require several $100$~mT of field to be resolved. 

The final term in ${\cal H}_0$ describes the interaction of the \nv electron spin with the \NN~nuclear spin:
\begin{equation}
\label{eq_HSI}
    {\cal H_{\rm SI}} =A_{\rm ||}\,S_{\rm z}\,I_{\rm z}+A_{\rm \perp}(S_{\rm x}\,I_{\rm x}+S_{\rm y}\,I_{\rm y}),
\end{equation}
where $A_{\rm ||}=2.14$~MHz and $A_{\rm \perp}=2.7$~MHz are the parallel and perpendicular components of the hyperfine tensor with respect to the \nv center axis~\cite{PhysRevB.79.075203}.

\begin{figure*}
	\centering 
		\includegraphics[width=\textwidth]{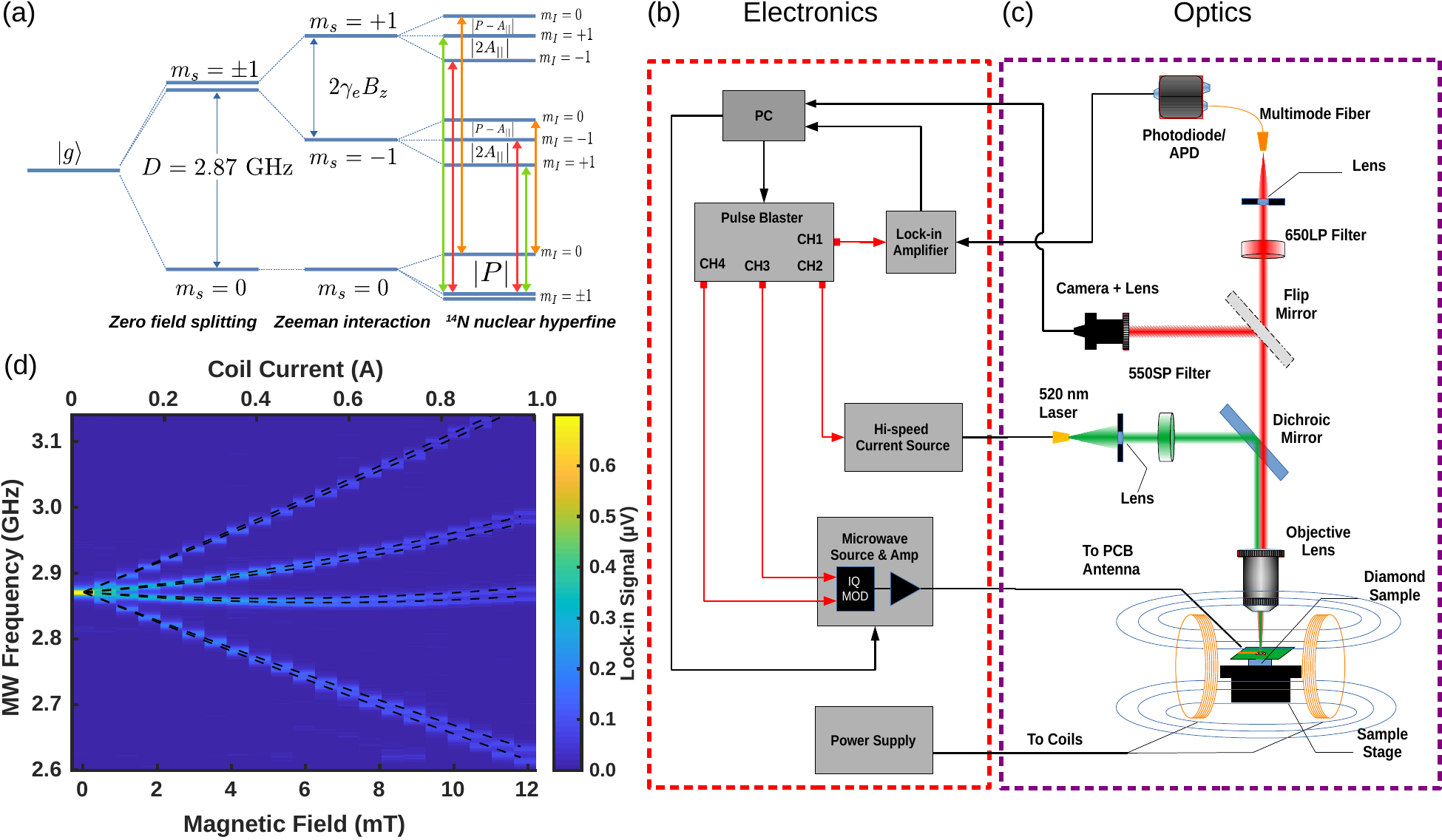}
		\caption{
		(a) Ground state energy levels based on the Hamiltonian in Eq.~\ref{eq_H0}, showing the zero field splitting, Zeeman splitting and \NN~nuclear hyperfine interaction. The differently colored arrows represent the allowed electron spin transitions for the different nuclear spin orientations $m_I=-1$ (red), $m_I=0$ (orange) and $m_I=+1$ (yellow). 
		(b) Electronics setup and (c) optics setup, identical to Ref.~\onlinecite{Sewani2020} except for the addition of electromagnetic coils and programmable power supply, used to create a well-defined $B_0$ field indicated by the blue field lines. 
		(d) ODMR spectra recorded for different magnetic fields generated by the electromagnets, measured using the photodiode. Each column of the 2D map corresponds to a single ODMR scan for a certain magnetic field strength. The electron spin resonance transitions split with increased magnetic field strength due to the Zeeman effect as described by Eq.~\ref{eq_HS}. The eight different transitions arise from the four different orientations of the NV$^-$ center axis, of which two pairs of two have approximately the same angle with respect to the external magnetic field $B_0$ [see also Fig.~\ref{fig2}(a)]. The dashed lines are the calculated transition frequencies using Eq.~\ref{eq_HS}.}
		\label{fig1}
\end{figure*}

In Fig.~\ref{fig1}(a) we show a schematic of the energy levels of the system (not to scale) according to the Hamiltonian ${\cal H}_0$ in Eq.~\ref{eq_H0}. We start with the bare ground state energy level $|{g}\rangle$ of the \nv center on the left side. Turning on the zero-field splitting $D$ then lifts the degeneracy between the different $S=1$ levels, by raising the $m_s=\pm1$ levels by $D=2.87$~GHz above the $m_s=0$ level. Next, we include the electron Zeeman interaction as given by $\gamma_{\rm e}\,\mathbf{B_0}\cdot\mathbf{S}$. Assuming that $\mathbf{B_0}$ is applied along the z-axis, the Zeeman interaction raises the $m_s=+1$ level by $2\gamma_{\rm e}\,B_{\rm z}$ above the $m_s=-1$ level. Finally, we include the \NN~nuclear spin. The quadrupole component $P$ lifts the degeneracy between the different $I=1$ nuclear spin levels by raising the $m_I=0$ level by $5.01$~MHz above the $m_I=\pm1$ levels (see Eq.~\ref{eq_HI}), and the hyperfine interaction between electron spin and nuclear spin further splits the $m_I=+1$ and $m_I=-1$ levels by $|2A_{\rm ||}|$ for the $m_s=\pm1$ electron spin states (see Eq.~\ref{eq_HSI}). The nuclear spin Zeeman interaction as defined by $\gamma_{\rm n}\,\mathbf{B_0}\cdot\mathbf{I}$ in Eq.~\ref{eq_HI} is not included in this schematic as it is too small to be observed in our experiments.
 
\begin{figure}[ht!] 
	\centering 
		\includegraphics[width=\columnwidth]{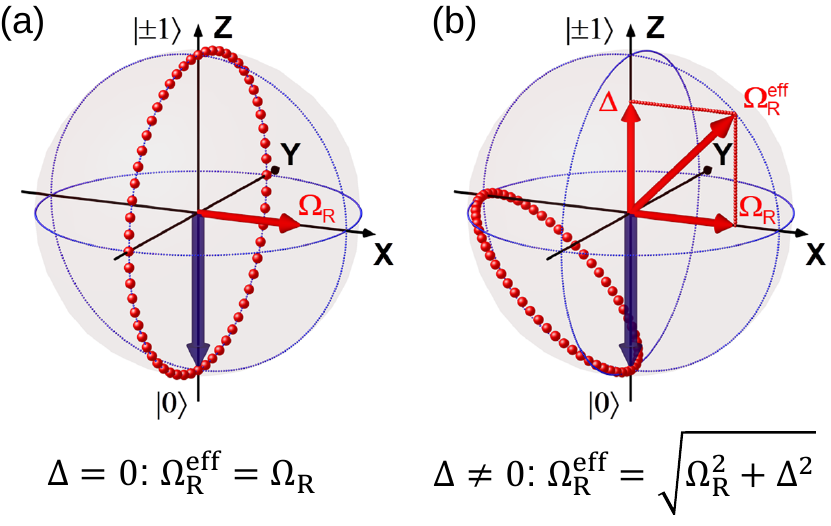}
		\caption{
		(a) Nutation of a spin around the Bloch sphere in the rotating frame, initialized in the $|{0}\rangle$ state and for a resonant drive (i.e., $\Delta=0$).  
		(b) Nutation of a spin around the Bloch sphere in the rotating frame, initialized in the $|{0}\rangle$ state and for a slightly detuned drive (i.e., $\Delta=\Omega_{\rm R}$).}
		\label{fig_bloch}
\end{figure}

\section{Rabi formula}\label{sec_RabiTheory}
One of the learning outcomes is the understanding of experimental non-idealities. A very common non-ideality is a driving field that is detuned in frequency from the spin resonance transition and that is strong in power. This will lead to non-perfect rotations of the spin, and the power-broadening of the ODMR resonance lines which prevents the observation of fine features of the spectrum at high driving powers. The effect of both can be easily understood via the Rabi formula, that describes the probability of flipping a spin (or any two-level system) as a function of pulse time when a coherent driving pulse is applied~\cite{Cohen1977}:
\begin{equation}
P(t) = \frac{\Omega_{\rm R}^2}{\Omega_{\rm R}^2 + \Delta^2}\,{\rm sin}^2\left(\frac{1}{2}\sqrt{\Omega_{\rm R}^2+\Delta^2}\,2\pi\,t  \right),
    \label{eq_Rabi}
\end{equation}
where $t$ is the length of the driving pulse, $\Delta = \nu_{\rm MW} - \nu_{\rm res}$ is the frequency detuning between the frequency of the driving field $\nu_{\rm MW}$ and the resonance frequency of the spin transition $\nu_{\rm res}$, and $\Omega_{\rm R}$ is the rotation frequency of the spin while it is driven -- the so-called Rabi frequency.\footnote{$\Delta$ and $\Omega_{\rm R}$ are defined in units of Hz.} In general $\Omega_{\rm R}=\tfrac{1}{\sqrt{2}}\gamma_e\,B_1$ for a spin-1 system, where $B_1$ is the amplitude of the oscillating magnetic field provided by an antenna~\cite{Vallabhapurapu2021}. Note, that this is different to a spin-1/2 system for which $\Omega_{\rm R}=\tfrac{1}{2}\gamma_e\,B_1$.

Eq.~\ref{eq_Rabi} can be considered in two parts. The first part is the envelope, $\frac{\Omega_{\rm R}^2}{\Omega_{\rm R}^2 + \Delta^2}$ that describes the amplitude of the Rabi oscillations. If the amplitude is plotted as a function of the frequency detuning $\Delta$, the result is a Lorentzian lineshape that has a half-width at half-maximum (HWHM) of $\Delta_{\rm HWHM}=\Omega_{\rm R}$. This means that for a detuning of $\Delta=\Omega_{\rm R}$ the spin can still be rotated with 50\% amplitude, clearly showing how a high Rabi frequency caused by a large $B_1$ amplitude leads to a large linewidth in the ODMR spectrum.

The second part of Eq.~\ref{eq_Rabi} is the time-dependent part ${\rm sin}^2\left(\frac{1}{2}\sqrt{\Omega_{\rm R}^2+\Delta^2}\,2\pi\,t \right)$ that describes the actual time evolution of the spin during the Rabi oscillations. The frequency of the Rabi oscillations is given by $\Omega_{\rm R}^{\rm eff}=\sqrt{\Omega_{\rm R}^2+\Delta^2}$\hspace{6pt}~\footnote{The cancellation of the $\tfrac{1}{2}$-term derives from the ${\rm sin}^2$-term. This is because a ${\rm sin}^2$-function oscillates at twice the frequency as the corresponding ${\rm sin}$-function.}, which reduces to $\Omega_{\rm R}^{\rm eff}=\Omega_{\rm R}$ for the resonant case with $\Delta = 0$. A profound consequence on experiments being performed at a frequency detuning $\Delta \neq 0$ is that the Rabi oscillations are faster and have a smaller amplitude than in the resonant case. This can be visualized by imagining a state on the Bloch sphere that starts at the south pole and rotates about different axes (see Fig.~\ref{fig_bloch}). For a rotation axis along the equator of the Bloch sphere, the path that the state would describe on the Bloch sphere would take it all the way to the north pole (i.e., large amplitude) as seen in Fig.~\ref{fig_bloch}(a). However, when the rotation axis cants closer to one of the poles, the path that the state describes on the Bloch sphere would not take it very far away from the south pole (i.e., small amplitude) as seen in Fig.~\ref{fig_bloch}(b).

\section{Equipment}\label{sec_Equip}
The experimental setup used for the experiments in this manuscript is almost identical to the one introduced in detail in Sec.~IV of Ref.~\onlinecite{Sewani2020}. 
A few improvements have been introduced to enable the observation of nuclear interactions in the experiments. Firstly, while a photodiode is sufficient to perform the measurements presented here, and was in fact used for Fig.~\ref{fig1}(d), the signal to noise ratio can be improved using an avalanche photodiode (APD), e.g., \textit{Excelitas Technologies SPCM-AQRH}. As with the photodiode, the APD can simply be connected to the lock-in amplifier input (in voltage mode), where the low-pass filter will average over the short voltage pulses. When these experiments were implemented in the teaching labs at UNSW Sydney, the students performed all experiments using standard photodiodes.

Secondly, as shown in Fig.~\ref{fig1}(c), we have added a pair of homemade electromagnetic coils, powered by a DC lab power supply, to generate a well-defined and homogeneous $B_0$ field. Each coil is constructed from a reel and an L-shaped mount (see Fig.~\ref{electromagnets}). A hole at the center of the reel, allows the addition of a core, in our case a supermendur cylinder, to further increase the magnetic field. The advantage over permanent magnets is that the magnetic field strength can be easily tuned by adjusting the current going through the coils, and the magnetic field direction is known as well. In App.~\hyperref[AppA]{A}, we present more details on the design and geometry of the homemade electromagnets. A low-cost, commercial solution are the \textit{Adafruit Industries 3875} electromagnets (see e.g. \textit{Digi-Key 1528-2691-ND}. 
An alternative way to measure a more homogeneous ensemble of \nv centers would be to interchange the multi-mode fiber for detection with a single-mode fiber to collect signal from a smaller sample volume. 

The sample used for the experiments presented here is a commercially available, off-the-shelf, $\langle100\rangle$-oriented CVD diamond sample from \emph{Element Six}~\cite{cvd}, and is used without any processing. The dimensions of the sample are $2.6\times2.6\times0.25$~mm, and the nitrogen concentration is $<1$~ppm.
In Appendix~C of Ref.~\onlinecite{Sewani2020}, there is a comparison between this sample and a high-pressure, high-temperature (HPHT) diamond sample that was used for the experiments in Ref.~\onlinecite{Sewani2020}. The CVD sample has a $\sim1000$ times lower PL signal, and while the spin signal decreases proportionally, we obtain high-quality data with only $\sim10$ times longer measurement times. In contrast, the CVD sample has a much longer coherence time than the HPHT sample, on account of the lower N concentration, that allows the observation of effects originating from coupling to nuclear spins. More recently, we have also started using a commercially available sample from \emph{Element Six} with an NV center density of $\sim 300$~ppb, optimized for quantum experiments~\cite{DNV}. This sample has a $>50\times$ higher ODMR signal than the CVD sample~\cite{cvd}, and was used for the experiments in Fig.~\ref{fig2}(a).

\section{Experiments}\label{sec_Exp}
All experiments described here are based on electron spin resonance (ESR), that allows us to coherently control the electron spin of the \nv center and perform different types of measurements depending on the exact pulse sequence. More precisely, the type of ESR experiment that we perform here is termed optically-detected magnetic resonance (ODMR), since the electron spin state is read out (and initialized) via optical means, i.e., laser excitation and spin-dependent photon emission via photoluminescence (PL). The exact process was already described in Ref.~\onlinecite{Sewani2020} and many other publications~\cite{Doherty2013,Zhang2018,Bucher2019,Misonou2020}.

As introduced in Ref.~\onlinecite{Sewani2020} and can be seen in Fig.~\ref{fig1}(a), there are a number of different ESR transitions that can be driven with an oscillating or rotating magnetic field $B_1$. For $B_0=0$~T and neglecting the nuclear spin, all transitions are degenerate and centered around $D=2.87$~GHz. When the magnetic field is turned on, the Zeeman splitting (Eq.~\ref{eq_HI}) lifts this degeneracy and leads to different resonance frequencies for $m_s=0\leftrightarrow m_s=-1$ and $m_s=0\leftrightarrow m_s=+1$. When the interaction with the \NN~nuclear spin is included, there are six possible transitions -- two each for $m_I=-1$ (red arrows), $m_I=0$ (orange arrows), and $m_I=-1$ (yellow arrows). These transitions will keep the nuclear spin state unchanged, but modify the electron spin state.

\subsection{Magnetic field dependence}\label{sec_Bfield}
\begin{figure}[ht!] 
	\centering 
		\includegraphics[width=\columnwidth]{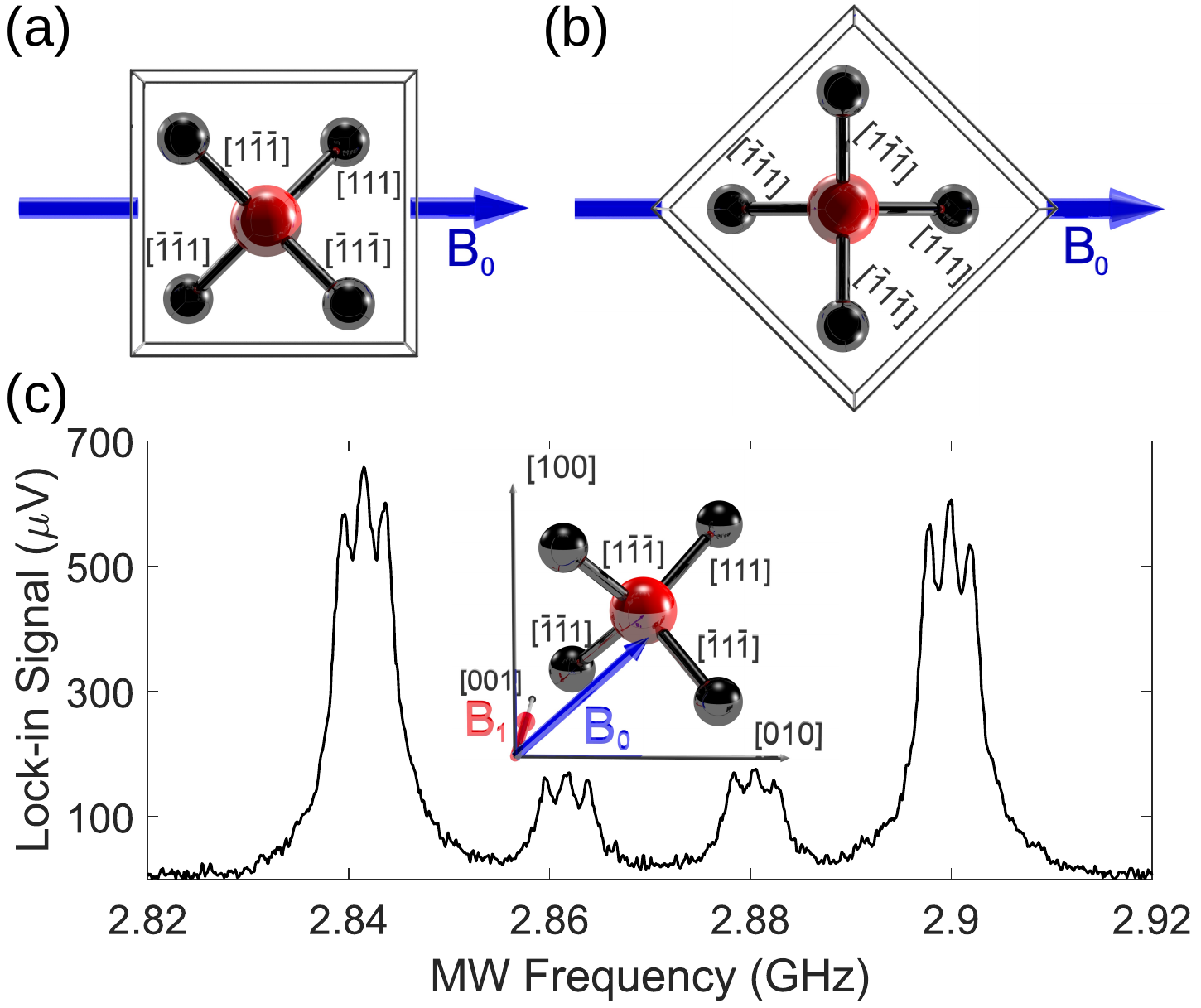}
		\caption{
		(a) Orientation of the NV crystallographic axes with respect to the sample facets and the external magnetic field $B_0$ when the sample is mounted in the square orientation (\Square).
		(b) Same as (a) with the sample mounted in the diamond orientation (\BigDiamondshape).
		(c) ODMR spectrum at a low magnetic field of $B_0=1.3$~mT, showing 12 observable hyperfine peaks. Here, the peaks of the $[111]$ and $[\bar{1}\bar{1}1]$ \nv orientations, and those of the $[1\bar{1}\bar{1}]$ and $[\bar{1}1\bar{1}]$ \nv orientations overlap, respectively. The inset shows the schematic of the alignment of the crystallographic axes with respect to the applied $B_0$ and $B_1$ magnetic fields. The $B_0$ field is oriented along the $[0.67, 0.74,0.02]$ direction. The red sphere indicates the N atom of the \nv center, while the black spheres indicate the four possible lattice sites for the vacancy. In this orientation, $B_0$ results in similar Zeeman splittings for the $[111]$ and $[\bar{1}\bar{1}1]$, and the $[1\bar{1}\bar{1}]$ and $[\bar{1}1\bar{1}]$ \nv orientations, respectively, while the orientation of $B_1$ results in similar Rabi frequencies $\Omega_{\rm R}$.}
		\label{figOrientation}
\end{figure}

We start by measuring the ODMR spectrum of the diamond sample as a function of magnetic field $B_0$. The faces of the CVD diamond chip are oriented along the $\langle100\rangle$ directions and the four orientations of \nv centers are along the [111], [1$\bar{1}\bar{1}$], [$\bar{1}1\bar{1}$], and [$\bar{1}\bar{1}1$] directions (see Fig.~\ref{figOrientation}). The $B_1$ field created by the printed-circuit-board antenna is applied along the [001] direction, and therefore perpendicular to the sample surface. The $B_0$ field created by the coils of the electromagnet is applied parallel to the sample surface. By rotating the CVD sample on the sample holder, different $B_0$ directions with respect to the crystallographic axes can be created. More specifically, when the sample is mounted in a square orientation [i.e., like \Square, as shown in Fig.~\ref{figOrientation}(a)] the $B_0$ field acts along the [100] direction. This results in all four \nv orientations experiencing the same effective angle between their \nv axes and $B_0$ and having the same two ODMR frequencies. However, when the sample is mounted in a diamond orientation [i.e., like \BigDiamondshape, as shown in Fig.~\ref{figOrientation}(b)] the $B_0$ field acts along the [110] direction, resulting in the same effective angle for the [111] and [$\bar{1}\bar{1}1$], and the [1$\bar{1}\bar{1}$] and [$\bar{1}1\bar{1}$] orientations, respectively, and a total of four ODMR frequencies. In order to lift the degeneracy of all four \nv orientations and see eight different transitions appear, $B_0$ needs to be additionally rotated out of the sample plane.

We mount our sample roughly in the diamond orientation (i.e., like \BigDiamondshape), i.e., with $B_0$ directed along the [110] direction, however, we introduce a slight misalignment and let the students work out the exact orientation of the magnetic field. They can achieve this by recording $B_0$-dependent ODMR spectra and mapping the observed splittings onto the calculated transition energies from Hamiltonian ${\cal H_{\rm S}}$ (Eq.~\ref{eq_HS}) using Matlab.

We plot such a magnetic field map in Fig.~\ref{fig1}(d). Here, the magnitude of $B_0$ can be increased without altering the magnetic field direction by increasing the current through the electromagnet (see top x-axis) from 0~A to 1.0~A in 21 steps. For each $B_0$ magnitude an ODMR spectra was recorded to build up the 2D ODMR map. The students then fit the simple electron spin Hamiltonian ${\cal H_{\rm S}}$ to the data to work out the coil constant that describes the relationship between current and $B_0$ field. The dashed lines in Fig.~\ref{fig1}(d) show the best fit for this data. The fit indicates that $B_0$ is applied along the [0.56 0.83 0.03] direction and the coil constant is 11.9~mT/A. 

Note that the resistance of the electromagnetic coils is about 15~$\Omega$ and that it increases when the coils heat up at larger driving currents. It is therefore important to use a current source to control the magnetic field. Furthermore, the coils will also heat up the diamond sample and therefore the \nv center environment. At a high current, the ODMR spectrum will shift according to the temperature dependency of the zero-field-splitting $dD/dT=-74.2$~kHz/K~\cite{Acosta2010}. For our experimental setup a current of 1~A heats up the diamond sample by $\sim40$~K and shifts the zero-field splitting by $\sim3$~MHz.

In the teaching labs at UNSW Sydney, the students recorded eleven ODMR spectra for electromagnet currents between 0~A and 0.5~A in 0.05~A steps. Each spectrum was measured from 2.65~GHz to 3.09~GHz in 221 steps with a wait time per point of 600~ms using the photodiode. This resulted in a total measurement time of $\sim25$~minutes. While the measurements were running, the students were given a prepared Matlab script to examine the relationship between the coil constant and the magnetic field orientation. The students were then able to iteratively extract the relevant $B_0$ parameters by fitting their values to the measurement data, comfortably within the allocated time frame of the lab. This exercise addresses the first learning outcome in Section II.

\begin{figure*}[ht!] 
	\centering 
		\includegraphics[width=0.95\textwidth]{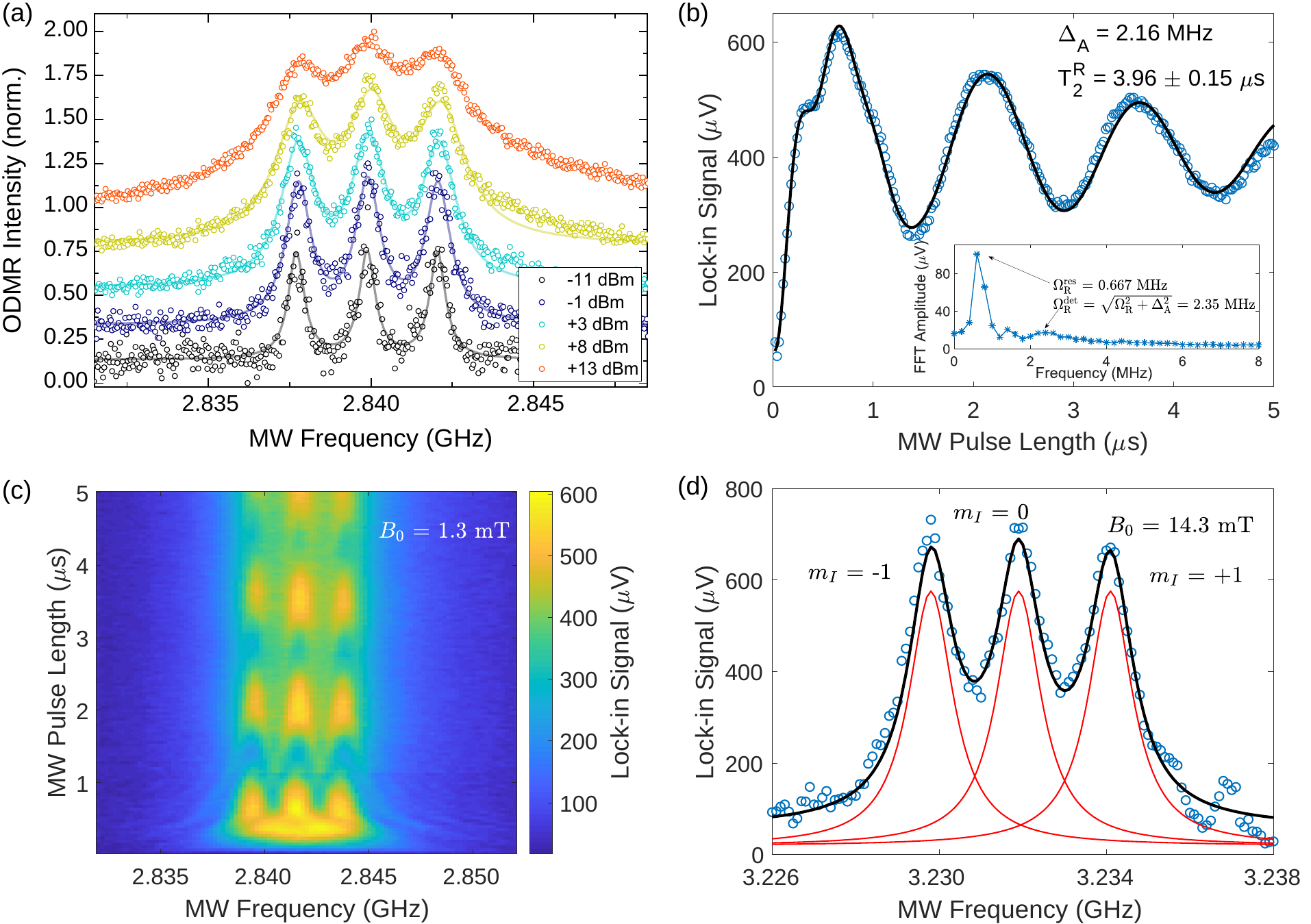}
		\caption{
		(a) ODMR spectra of the $m_s=0 \longleftrightarrow -1$ transition at a low magnetic field of $B_0=1.3$~mT and for different MW powers. The circles are experimental data points, while the solid lines are fits to three Lorentzian peaks with equal width and area. The curves are offset by multiples of 0.25 for better visibility. At high powers, the power broadening prevents us from clearly resolving the three hyperfine lines.
		(b) Rabi oscillations of the $m_s=0 \longleftrightarrow -1$ transition for $m_I=0$ at $B_0=1.3$~mT. The inset shows the corresponding frequency domain information. Highlighted are the resonant Rabi frequency $\Omega_{\rm R}^{\rm res}=0.667\pm0.005$~MHz for driving the $m_I=0$ hyperfine line, and the detuned Rabi frequency $\Omega_{\rm R}^{\rm det}=2.35\pm0.06$~MHz for driving the $m_I=\pm1$ hyperfine lines. 
		(c) Rabi chevron map of the $m_s=0 \longleftrightarrow -1$ transitions at $B_0=1.3$~mT, showing the PL signal as a function of MW frequency and MW pulse length. 		
		(d) Zoomed-in ODMR spectra of the $m_s=0 \longleftrightarrow +1$ transitions of the $[111]$ \nv orientation at $B_0=14.3$~mT. The three different nuclear spin states can be clearly resolved.}
		\label{fig2}
\end{figure*}

\subsection{Spectral signature of hyperfine coupled \texorpdfstring{\NN~}{14N}}\label{sec_Spec14N}
Having mapped out the Zeeman splitting of the $S=1$ \nv center electron spin in Fig.~\ref{fig1}(d), we can now take a closer look at the ODMR spectrum to examine the spectral signature of the hyperfine coupled \NN~nuclear spins. Owing to the optical properties of the \nv center, the electron-nuclear spin interactions for the native \NN~nucleus can be seen directly in the ODMR spectrum. The \NN~hyperfine interaction tensor (see Eq.~\ref{eq_HSI}) splits each electron energy level [see Fig.~\ref{fig1}(a)], causing the electron spin transitions to become nuclear spin dependent, with a $2.16$~MHz splitting in between the different $I_{z}$ projections. The different colored arrows in Fig.~\ref{fig1}(a) represent the allowed electron spin transitions for the different nuclear spin orientations $m_I=-1$ (yellow), $m_I=0$ (orange) and $m_I=+1$ (red). These splittings can be resolved with a high-quality, low concentration \nv center ensemble sample ($\leq1$ppm~\cite{van1997dependences}), a high ODMR frequency resolution and a low microwave power~\cite{Smeltzer_2011}.

In Fig.~\ref{figOrientation}(c) we show the ODMR spectrum at a magnetic field of $B_0=1.3$~mT along the [0.67, 0.74, 0.02] direction (as estimated from the spectrum). The field is only slightly misaligned from the [110] crystallographic axis [see also Fig.~\ref{figOrientation}(c) -- inset], leading to similar Zeeman splittings and overlapping transitions for the $[111]$ and $[\bar{1}\bar{1}1]$, and the $[1\bar{1}\bar{1}]$ and $[\bar{1}1\bar{1}]$ \nv orientations, respectively. All the coherent control measurement results in the following sections (i.e., Figs.~\ref{fig2}, \ref{fig3}, \ref{fig4}) are taken based on this field strength and orientation. 
Each peak in Fig.~\ref{figOrientation}(c) displays a triplet structure that corresponds to the three different nuclear spin orientations. In the following sections, we will examine closely what effect the presence of the hyperfine lines has in coherent control experiments such as Rabi oscillations, Ramsey oscillations and Hahn echoes.

From the Rabi formula (see Section~\ref{sec_RabiTheory}), we know that a lower microwave power results in a narrower HWHM (Half Width at Half Maximum) of the peaks in the spectrum. We map out this dependency in Fig.~\ref{fig2}(a), where we plot five ODMR spectra recorded for different MW powers (open circles), together with a fit to the sum of three Lorentzian peaks with equal width and area (solid lines). At the lowest MW power, the three hyperfine peaks are clearly separated, while the power broadening at the higher powers leads to an overlap of the individual peaks.  

We therefore need to work with low MW powers if we want to resolve the hyperfine splitting without power broadening. This requires a balance between resolution and signal strength, and we noticed that it is advantageous to work at low magnetic fields where the quenching of the photoluminescence is insignificant and the electron spin transitions from two different NV$^-$ orientations with slightly different angles overlap. For all following experiments, however, we choose a MW power that leads to some overlap between peaks [similar to the $+3$~dBm curve in Fig.~\ref{fig2}(a)] in order to observe effects of simultaneously driving neighboring transitions.

In principle, it is also possible to work with NV$^-$ centres of a single orientation. To achieve this, a slightly misaligned and higher magnetic field is needed to split the electron spin transitions associated with the four different NV$^-$ orientations.
\footnote{As this results in a lower photoluminescence signal, we use continuous ODMR to compensate for the decrease of signal strength, where the laser is on throughout the whole lock-in cycle, while the MW is pulsed on only during the first half.}
Fig.~\ref{fig2}(d) shows an example spectrum of the electron spin transitions between $m_s=0$ and $m_s=+1$ at $B_0=14.3$~mT. The three different nuclear spin states can be clearly resolved and are labeled accordingly. We fit the spectrum with the sum of three Lorentzian peaks with equal amplitude and FWHM shown as black line, with the individual Lorentzian peaks show as red lines.

\subsection{Rabi oscillations in the presence of \texorpdfstring{$^{14}$N}{14N}}\label{sec_Rabi}

Applying a microwave pulse in resonance with one of the three hyperfine lines, caused by the $^{14}$N nuclear spin, induces coherent electron spin oscillations between the $m_s=0$ and $m_s=\pm 1$ spin eigenstates, known as Rabi oscillations.
However, it is important to realize that the microwaves are simultaneously driving the detuned hyperfine transitions, due to the finite linewidth of the individual lines as well as the power broadening as described by the Rabi formula [e.g. see overlap of peaks in Fig.~\ref{fig2}(a,c,d)].

Figure~\ref{fig2}(b) shows a Rabi oscillation experiment, driven with the microwave frequency in resonance with the $m_I=0$ transition [middle peak of the left triplet in Fig.~\ref{fig2}(a)], and the data is fitted to the sum of two exponentially decaying sinusoids $A_\alpha\sin(2\pi f_\alpha) \exp(-\tau/T_\alpha)+A_\beta\sin(2\pi f_\beta) \exp(-\tau/T_\beta)$ (black line). The frequencies for the two sinusoids correspond to the on-resonance drive of the $m_I=0$ transition and the detuned drive of the $m_I=\pm1$ transitions. The two oscillation frequencies become apparent when taking the fast Fourier transform (FFT) of the Rabi oscillation experiment data [see Fig.~\ref{fig2}(b) -- inset], showing peaks at $\Omega_{\rm R}^{\rm res}\approx0.7$~MHz and $\Omega_{\rm R}^{\rm det}\approx2.3$~MHz. 

According to the Rabi formula (Eq.~\ref{eq_Rabi}), the Rabi oscillation frequency is given by:
\begin{equation}
    \Omega_{\rm R}^{\rm eff}=\sqrt{(\Omega_{\rm R}^{\rm res})^2+\Delta^2},
    \label{eq2}      
\end{equation}
where $\Omega_{\rm R}^{\rm res}$ is the on-resonance Rabi oscillation frequency and $\Delta$ is the detuning. $\Omega_{\rm R}^{\rm res}$ is given by:
\begin{equation}
    \Omega_{\rm R}^{\rm res}=\tfrac{1}{\sqrt{2}}\gamma{}_{\rm e}B_{1},
\end{equation}
where $B_1$ is the oscillating magnetic field component perpendicular to the NV$^-$ axis delivered by the antenna~\cite{Pham2013MagneticFS}. Therefore, the Rabi frequency of $\Omega_{\rm R}=0.667\pm0.005$~MHz in Fig.~\ref{fig2}(b) corresponds to a field amplitude of $B_1=0.034$~mT, and we keep it fixed to this value for all coherent control experiments.
The faster frequency of $\Omega_{\rm R}^{\rm det}=2.35\pm0.06$~MHz in Fig.~\ref{fig2}(b) corresponds to Rabi oscillations on the detuned hyperfine transition and agrees well with the one calculated using Eq.~\ref{eq2}, where $\Delta_A=2.16$~MHz is the detuning between adjacent hyperfine peaks: $\Omega_{\rm R}^{\rm det}=\sqrt{(\Omega_{\rm R}^{\rm res})^2+\Delta_A^2}=\sqrt{(0.667~{\rm MHz})^2+(2.16~{\rm MHz})^2}=2.26$~MHz. The Rabi coherence time extracted from the on-resonance component is $T_2^{\rm R}=3.96$~$\mu$s, which is most likely limited by the inhomogeneity of the $B_1$ field over the sample area~\cite{Vallabhapurapu2021}.

The effect of power broadening and cross-talk between the different hyperfine lines can be well visualized in a so-called Rabi chevron experiment, where the MW frequency and the pulse length are both swept -- or in other words, a Rabi oscillation measurement is performed for various MW frequencies. Note that due the large number of data points, a Rabi chevron is usually outside of the scope of a teaching lab.
Fig.~\ref{fig2}(c), shows such a measurement for the same triplet of peaks as in Fig.~\ref{fig2}(a,b). Each of the three hyperfine lines shows its own chevron fringes, overlapping with the neighbouring lines, and leading to the deviation from a single, decaying sinusoid in Fig.~\ref{fig2}(b). 

In principle, a simultaneous drive of all three hyperfine lines with similar strength can be achieved by applying a MW drive with stronger power to increase the Rabi oscillation frequency beyond the hyperfine coupling ($2.16$~MHz)~\cite{Vallabhapurapu2021}. However, in our experiments the MW power was purposefully kept low, so that the effect of hyperfine transitions can be observed. This allows the students to develop an intuitive understanding of the consequences and origin of power broadening, fulfilling the second learning outcome in section II.

The Rabi experiment discussed in this section is the gateway between performing spin resonance spectroscopy and applying discrete gate operations in pulse sequences. Extracting the Rabi frequency $\Omega_{\rm R}=0.667$~MHz allows us to calibrate the lengths of our MW pulses to perform controlled rotations around the Bloch sphere. Here a $\pi$-pulse corresponds to a MW pulse of correct length $\tau_\pi=0.75$~$\mu$s to rotate the electron spin $180^\circ$ around the Bloch sphere, e.g. from $m_s=0$ to $m_s=-1$. A $\pi/2$-pulse of length $\tau_{\pi/2}=0.38$~$\mu$s rotates the electron spin by $90^\circ$, e.g. from the $m_s=0$ state to the $|{+y}\rangle=\tfrac{1}{\sqrt{2}}({|0\rangle}+{i|{-1}\rangle})$ superposition state at the equator of the Bloch sphere. Once the pulse lengths have been calibrated, experiments using pulse sequences, such as the coherence times measurements in the next sections can be implemented. See also Ref.~\onlinecite{Sewani2020} for a more detailed discussion and visualizations on the spin rotations.

\begin{figure} 
	\centering 
		\includegraphics[width=0.9\columnwidth]{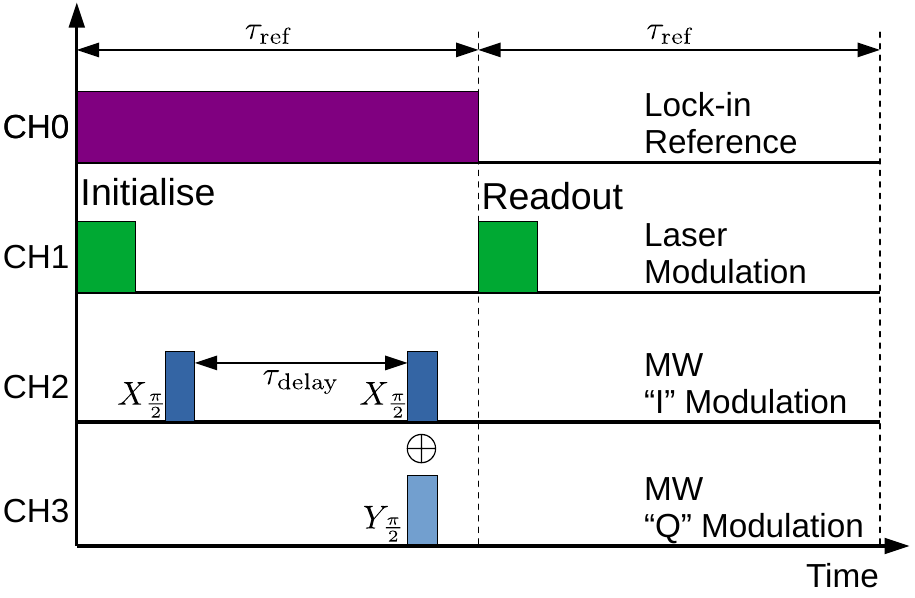}
		\caption{Ramsey pulse sequence used for the measurements in Fig.~\ref{fig3}. Channel 0 (CH0) provides the reference for the lock-in amplifier, channel 1 (CH1) modulates the laser output for the optical initialization and readout pulses, channel 2 (CH2) modulates the ``I'' output, and channel 3 (CH3) modulates the ``Q'' output of the MW IQ modulator. The second $\pi/2$-pulse is either applied along the $X$-axis to implement the standard Ramsey sequence $X_{\pi/2} - \tau - X_{\pi/2}$ or about the $Y$-axis to implement the modified Ramsey sequence $X_{\pi/2} - \tau - Y_{\pi/2}$.}
		\label{fig-ram-ps}
\end{figure}

\subsection{Ramsey fringes in the presence of \texorpdfstring{$^{14}$N}{14N}}\label{sec_Ram}
\begin{figure*} 
	\centering 
		\includegraphics[width=\textwidth]{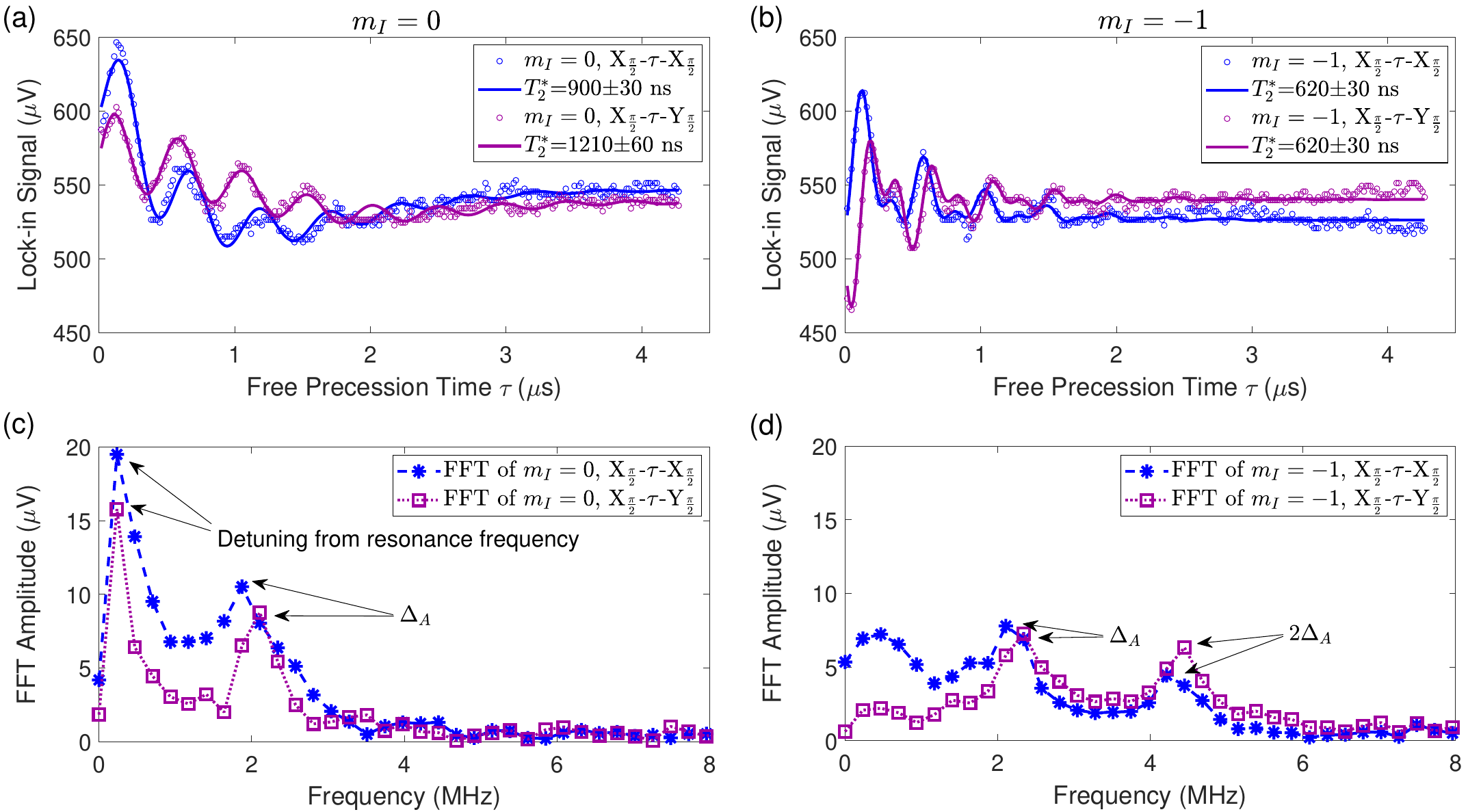}
		\caption{
		(a) Ramsey fringes measured on the $m_s=0 \longleftrightarrow -1$ transition for $m_I=0$ at $B_0=1.3$~mT. The blue circles correspond to the standard Ramsey sequence $X_{\pi/2} - \tau - X_{\pi/2}$, while the purple circles correspond to the modified Ramsey sequence $X_{\pi/2} - \tau - Y_{\pi/2}$. The solid lines are fits to the data with $T_2^*=900\pm30$~ns and $T_2^*=1210\pm60$~ns for the two pulse sequences, respectively. 
		(b) Same as (a) measured for $m_I=-1$. The extracted coherence times are $T_2^*=620\pm30$~ns and $T_2^*=620\pm30$~ns for the two pulse sequences, respectively. 
		(c) Fourier transform of the Ramsey fringes in (a) for $m_I=0$. Highlighted are the detuning from the resonant $m_I=0$ hyperfine line, and the detuning $\Delta_{A}$ from the $m_I=\pm1$ hyperfine lines.
		(d) Fourier transform of the Ramsey fringes in (b) for $m_I=-1$. Highlighted are the detuning $\Delta_{A}$ from the $m_I=0$ hyperfine line, and the detuning $2\Delta_{A}$ from the $m_I=-1$ hyperfine line.
		}
		\label{fig3}
\end{figure*}

A Ramsey experiment measures the free-induction decay or $T_2^*$ coherence time of a spin or a spin ensemble. The $T_2^*$ time is affected not only by noise during the pulse sequence, but also by detunings between the transition and the MW frequency. This is because the sequence does not contain any refocussing or echo pulses, in contrast to the Hahn echo and Carr-Purcell-Meiboom-Gill (CPMG) sequences discussed in Section~\ref{sec_Hahn}. 

Figure~\ref{fig-ram-ps} shows the pulse sequence that we use in our experiments and the outputs of the individual channels (CH0 - CH3) of the Pulse Blaster. The spins are initialized into the ${|0\rangle}$ state by the first laser pulse and read out by the second laser pulse (see CH1). The MW pulse sequence (See CH2 and CH3) is applied only during the first half-cycle for the lock-in amplifier (see CH0), so that the lock-in detection scheme is sensitive to changes in the photoluminescence caused by the MW pulse sequence (see also Ref.~\onlinecite{Sewani2020}). The first microwave $\pi/2$-pulse, applied along the $X$-axis, rotates the NV$^-$ electron spins from $|0\rangle$ to the $Y$-axis of the Bloch sphere, corresponding to the ${|{+y}\rangle}=\tfrac{1}{\sqrt{2}}({|0\rangle}+i{|{-1}\rangle})$ superposition state in the rotating frame. 
We then let the spins dephase freely for a duration $\tau$, before applying a second microwave $\pi/2$-pulse along the $X$-axis to transfer the NV$^-$ spin polarization from ${|{+y}\rangle}$ to ${|{-1}\rangle}$ for readout. A free induction decay curve can be recorded by varying the free precession time $\tau$ and the resulting data shows an exponential decay in ${|{-1}\rangle}$ population as a function of $\tau$ with decay time $T_2^*$ caused by decoherence and inhomogeneities~\cite{Pham2013MagneticFS}. For NV$^-$, the main inhomogeneities come from the variations in local strain and electric fields, as well as temporal and spatial fluctuations of the surrounding $^{13}$C nuclear spin bath~\cite{Doherty2013}. In addition, a constant frequency detuning between the spins and the MW drive will lead to a sinusoidal modulation of the data. This is because the spins precess at the speed of the detuning in the reference frame of the MW source, causing them to rotate from ${|{+y}\rangle}$ towards ${|{+x}\rangle}$ or ${|{-x}\rangle}$ and onward, depending on the sign of the detuning. As the second $\pi/2$-pulse in the sequence maps ${|{+y}\rangle}\rightarrow{|{-1}\rangle}$, ${|{+x}\rangle}\rightarrow{|{+x}\rangle}$, ${|{-y}\rangle}\rightarrow{|{0}\rangle}$, and ${|{-x}\rangle}\rightarrow{|{-x}\rangle}$, the constant detuning manifests as a sinusoidal oscillation in the ${|{-1}\rangle}$ population at the end of the sequence. As a consequence, a Ramsey sequence is often used for the accurate estimation of shifts in the spin qubit resonance frequency, as is important in magnetometry experiments~\cite{Barry2020}.   

In Fig.~\ref{fig3}(a) we show in blue the Ramsey data obtained for driving the NV$^-$ in resonance with the $m_I=0$ transition [same as the Rabi oscillations in Fig.~\ref{fig2}(b)]. The data shows the exponential decay superimposed with a sinusoidal modulation. We fit the data to the function $\exp(-\tau/T_2^*)\sum_{i=1}^{n}[A_i\sin(2\pi f_{i})]$, where $n$ corresponds to the number of expected transitions~\cite{Grezes2016}. Looking at the corresponding FFT of the data in Fig.~\ref{fig3}(c), we can clearly see two peaks. The peak at low frequencies indicates a slight detuning of our MW source to the $m_I=0$ transition frequency. The smaller peak at $\sim2$~MHz corresponds to the signal from the detuned $m_I=\pm1$ hyperfine lines, which are detuned from the $m_I=0$ transition by $\Delta_A=2.16$~MHz. 

In Fig.~\ref{fig3}(b), we show very similar data with the Ramsey experiment performed in resonance with the $m_I=-1$ transition. The data now shows two discrete modulation frequencies, one for the $m_I=0$ electron spin subsystem that is detuned by $\Delta_A$ and a second one for the $m_I=+1$ subsystem that is detuned by $2\Delta_A$. The two frequency components can be clearly seen as peaks in the Fourier-transform of the decay data in Fig.~\ref{fig3}(d).

For both Ramsey experiments in Figs.~\ref{fig3}(a) and (b) we also show data that is obtained by applying the second microwave $\pi/2$-pulse along the $Y$-axis (purple data). This modifies the mapping of the spin before readout to ${|y\rangle}\rightarrow{|{y}\rangle}$, ${|x\rangle}\rightarrow{|{0}\rangle}$, ${|{-y}\rangle}\rightarrow{|{-y}\rangle}$, and ${|{-x}\rangle}\rightarrow{|{-1}\rangle}$. As the projection pulse along the $Y$-axis (purple data) is shifted by $\pi/2$ in phase with respect to the projection pulse along the $X$-axis (blue date), the sinusoidal modulations are also shifted in phase by $\pi/2$ as seen in Figs.~\ref{fig3}(a) and (b). The possibility to modify the projection axis by adjusting the phase of the MW source forms the basis for quantum state tomography~\cite{Nielsen2002} and allows the extension of the teaching labs to include further experiments.

\subsection{Hahn echo in the presence of \texorpdfstring{\CC~}{13C}}\label{sec_Hahn}
\begin{figure} 
	\centering 
		\includegraphics[width=\columnwidth]{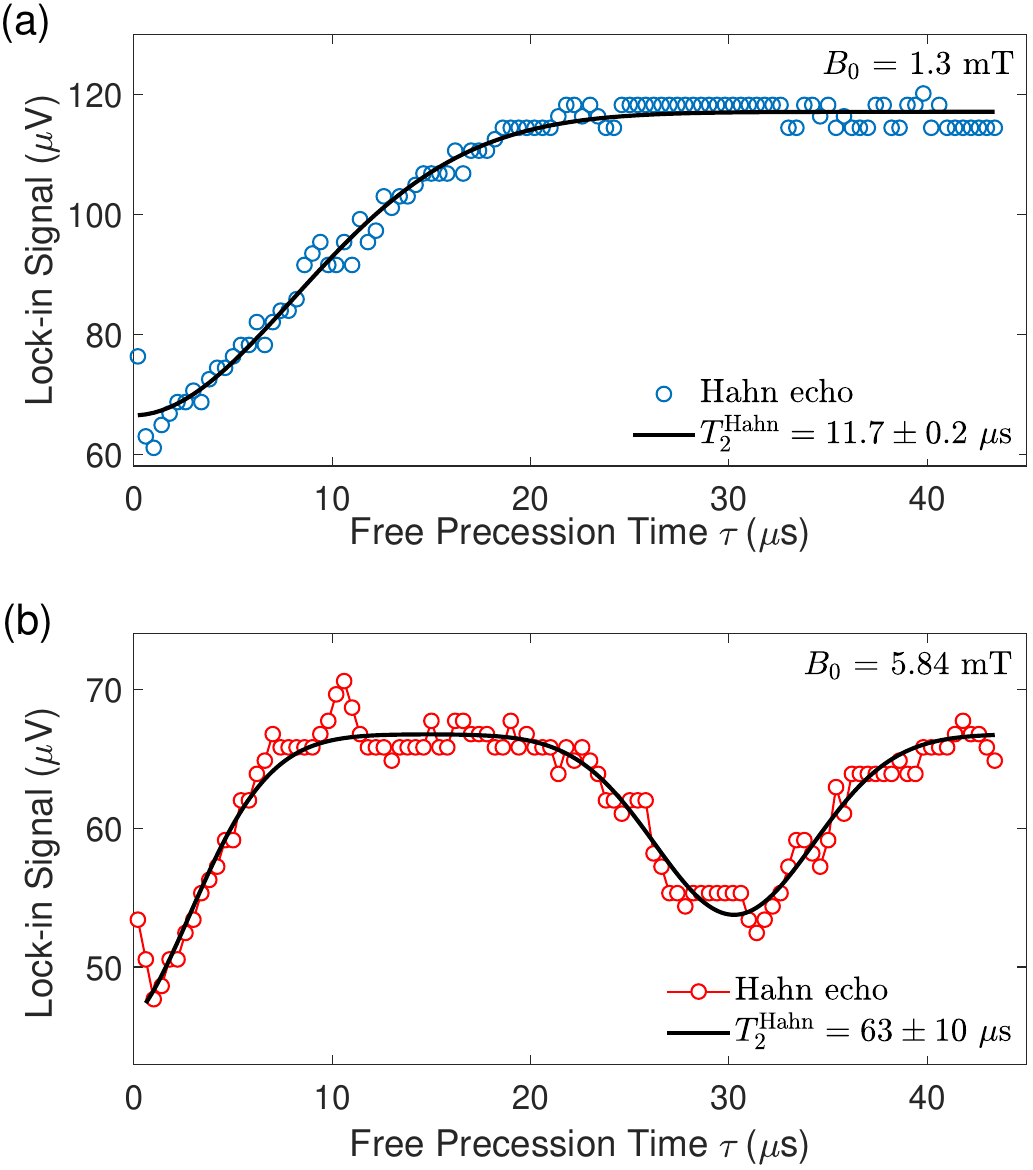}
		\caption{
		(a) Hahn echo decay measured on the $m_s=0 \longleftrightarrow -1$ transition for $m_I=0$ at a low magnetic field of $B_0=1.3$~mT. From a fit to the decay we extract $T_2^{\rm Hahn}=11.7\pm0.2$~$\mu$s.
		(b) Hahn echo decay measured at a higher magnetic field of $B_0=5.84$~mT. Coherence revival from the Larmor precession of the $^{13}$C nuclear spins can be seen at $\tau=30$~$\mu$s. From a fit to the data (black line), we extract $T_2^{\rm Hahn}=63\pm10$~$\mu$s.
		}
		\label{fig4}
\end{figure}

The Hahn echo experiment is designed to cancel out constant detunings and frequency offsets by adding an extra $\pi$-pulse to the sequence. The $\pi$-pulse inverts the spins halfway through the precession and as such refocusses variations in their Larmor precession frequency~\cite{Hahn1950}. 
We have already introduced this experiment in Ref.~\onlinecite{Sewani2020} for the purpose of dynamical decoupling, where it helps to improve the transverse relaxation time of the relevant spins, allowing the coherence time to be extended beyond the free induction decay time. 

The MW pulse sequence we implement for the Hahn echo measurement is $X_{\pi/2}-\tau-X_\pi-\tau-X_{\pi/2}$. In this sequence the electron spins are initialised in the $\ket{0}$ state and rotated into the superposition state $|y\rangle=\frac{1}{\sqrt{2}}(\ket{0} + i\ket{-1})$ using a MW $\frac{\pi}{2}$-pulse. The spins are then left to freely precess for a length of time $\tau$, after which they are inverted on the Bloch sphere with a $\pi$-pulse. The spins are again left to precess for the same amount of time, causing them to refocus before being rotated back into the $\ket{0}$ state with another $\frac{\pi}{2}$-pulse for optical readout. The spin signal is then measured as a function of the free precession time $\tau$, and the $T_{2}^{\rm Hahn}$ coherence time is given as the time constant of the exponential decay of the signal.

The Hahn echo can also be used to reveal interactions between spins -- in this case the hyperfine coupling to the \textsuperscript{13}C nuclei in proximity to the NV spins that have a natural abundance of $1.1$\%. The randomly distributed $^{13}$C nuclei of the diamond lattice produce a local magnetic field that is modulated by their Larmor precession frequencies, effectively resulting in an oscillating background magnetic field. Since the Hahn echo sequence can only remove slow fluctuations in detuning, the Larmor precession of the nuclei will lead to gradual decoherence of the electron spin. However, if the refocussing pulse occurs after exactly one period of nuclear spin precession, the background field modulation in both free precession intervals is exactly the same and the coherence reappears.

In Fig.~\ref{fig4}(a), we present the Hahn echo signal measured at a magnetic field of $B_0=1.3$~mT. From the exponential decay, we extract $T_{2}^{\rm Hahn}=11.7\pm0.2$~$\mu$s. Upon further investigation, a measurement at a higher magnetic field of $B_0=5.84$~mT reveals the presence of the above-mentioned coherence revival within the Hahn echo signal, as shown in Fig.~\ref{fig4}(b). 
The decay and revival in Fig.~\ref{fig4}(b) is attributed to the weak hyperfine coupling to the \textsuperscript{13}C spins near the NV center lattice sites~\cite{Childress281}. The signal revival at $2\tau\sim 30$~$\mu$s corresponds to a frequency of $\sim66.7$~kHz, which is consistent with the $f_{\rm C}=\gamma_{\rm C}B_0=62.5$~kHz Larmor precession frequency of the spin-1/2 nuclei with $\gamma_{\rm C}=10.705$~MHz/T and $B_0=5.84$~mT~\cite{Childress281,PhysRevB.82.201201}. 
The black line is a fit of the spin signal to two inverted Gaussian peaks whose amplitude follows an exponential decay $\propto\exp[-(2\tau/T_{2}^{\rm Hahn})]$, which gives $T_2^{\rm Hahn}=63\pm10$~$\mu$s. According to Ref.~\onlinecite{PhysRevB.82.201201}, the $T_2^{\rm Hahn}$ of an NV$^-$ ensemble can be as long as $600$~$\mu$s at room temperature. However, the coherence time reduces when the magnetic field is not parallel to the NV$^-$ axis, as is the case in our samples and experiments. 

As the coherence revival rate is linked to the Larmor precession frequency of the $^{13}$C nuclei and therefore proportional to the applied magnetic field, this revival is not present in Fig.~\ref{fig4}(a) where the lower magnetic field of $B_0=1.3$~mT would have led to a revival at $2\tau=144$~$\mu$s, longer than the $T_{2}^{\rm Hahn}$ coherence time.

\section{Conclusion}
Based on the cost effective, room temperature teaching setup described in Ref.~\onlinecite{Sewani2020} with some small improvements (electromagnet and lower NV$^-$ density diamond sample), we have presented here a set of more advanced experiments that aim to provide students with an in-depth understanding of physical concepts that are important for various quantum technologies.
The experiments are designed to highlight the interaction of the NV$^-$ centers with surrounding nuclear spins via the hyperfine interactions.  
The experiments allow the students to gain a better understanding of experimental imperfections as well as experience the manifestation of spin-spin interactions with the intrinsic $^{14}$N nuclear spin and the bath of surrounding $^{13}$C nuclear spins. These interactions become visible in the presented Rabi oscillation, free induction decay, and Hahn echo experiments. 

Despite the weaker signal intensity of the lower-density NV$^-$ diamond sample used here, the experiments work robustly on a standard desk in room light, and have partially been implemented in a fourth year undergraduate course at UNSW Sydney. Furthermore, the possibility for students to design their own pulse sequences and collect measurement data in real-time, makes this setup also an interesting option for more complex experiments in an undergraduate thesis. 

\begin{acknowledgments}
We would like to thank Jean-Philippe Tetienne (RMIT), Marcus Doherty (ANU) and Damian Kwiatkowski (TU Delft) for useful discussions. We acknowledge support from the School of Electrical Engineering and Telecommunications at UNSW Sydney, and the Australian Research Council (CE170100012). H.V. acknowledges support from the Sydney Quantum Academy. H.R.F. acknowledges the support of an Australian Government Research Training Program Scholarship. J.J.P. is supported by an Australian Research Council Discovery Early Career Research Award (DE190101397). A.L. and C.A. acknowledge support through the UNSW Scientia Program.
\end{acknowledgments}

\bibliography{mybib}

\begin{thebibliography}{10}
\expandafter\ifx\csname url\endcsname\relax
  \def\url#1{\texttt{#1}}\fi
\expandafter\ifx\csname urlprefix\endcsname\relax\def\urlprefix{URL }\fi
\providecommand{\bibinfo}[2]{#2}
\providecommand{\eprint}[2][]{\url{#2}}

\bibitem{Sewani2020}
\bibinfo{author}{Sewani, V.~K.} \emph{et~al.}
\newblock \bibinfo{title}{{Coherent control of NV− centers in diamond in a
  quantum teaching lab}}.
\newblock \emph{\bibinfo{journal}{American Journal of Physics}}
  \textbf{\bibinfo{volume}{88}}, \bibinfo{pages}{1156--1169}
  (\bibinfo{year}{2020}).
\newblock \urlprefix\url{https://doi.org/10.1119/10.0001905}.

\bibitem{Omdia2019}
\bibinfo{title}{Quantum computing for enterprise markets}.
\newblock
  \bibinfo{howpublished}{\url{https://tractica.omdia.com/research/quantum-computing-for-enterprise-markets/}}
  (\bibinfo{year}{2019}).
\newblock \bibinfo{note}{Accessed: 2020-04-14}.

\bibitem{QCReport2020}
\bibinfo{title}{Quantum computing report -- jobs}.
\newblock
  \bibinfo{howpublished}{\url{https://quantumcomputingreport.com/jobs/}}
  (\bibinfo{year}{2020}).
\newblock \bibinfo{note}{Accessed: 2020-04-14}.

\bibitem{Doherty2013}
\bibinfo{author}{Doherty, M.~W.} \emph{et~al.}
\newblock \bibinfo{title}{The nitrogen-vacancy colour centre in diamond}.
\newblock \emph{\bibinfo{journal}{Physics Reports}}
  \textbf{\bibinfo{volume}{528}}, \bibinfo{pages}{1--45}
  (\bibinfo{year}{2013}).
\newblock \urlprefix\url{https://doi.org/10.1016/j.physrep.2013.02.001}.

\bibitem{Zhang2018}
\bibinfo{author}{Zhang, H.} \emph{et~al.}
\newblock \bibinfo{title}{Little bits of diamond: Optically detected magnetic
  resonance of nitrogen-vacancy centers}.
\newblock \emph{\bibinfo{journal}{American Journal of Physics}}
  \textbf{\bibinfo{volume}{86}}, \bibinfo{pages}{225--236}
  (\bibinfo{year}{2018}).
\newblock \urlprefix\url{https://doi.org/10.1119/1.5023389}.

\bibitem{Qutools}
\bibinfo{author}{Qutools}.
\newblock \bibinfo{title}{Quantum sensing by diamond magnetometer}.
\newblock \bibinfo{howpublished}{\url{https://www.qutools.com/qunv/}, accessed
  14/04/2020} (\bibinfo{year}{2020}).

\bibitem{Misonou2020}
\bibinfo{author}{Misonou, D.} \emph{et~al.}
\newblock \bibinfo{title}{Construction and operation of a tabletop system for
  nanoscale magnetometry with single nitrogen-vacancy centers in diamond}.
\newblock \emph{\bibinfo{journal}{AIP Advances}} \textbf{\bibinfo{volume}{10}},
  \bibinfo{pages}{025206} (\bibinfo{year}{2020}).
\newblock \urlprefix\url{https://doi.org/10.1063/1.5128716}.

\bibitem{Bucher2019}
\bibinfo{author}{Bucher, D.~B.} \emph{et~al.}
\newblock \bibinfo{title}{Quantum diamond spectrometer for nanoscale {NMR} and
  {ESR} spectroscopy}.
\newblock \emph{\bibinfo{journal}{Nature Protocols}}
  \textbf{\bibinfo{volume}{14}}, \bibinfo{pages}{2707--2747}
  (\bibinfo{year}{2019}).
\newblock \urlprefix\url{https://doi.org/10.1038/s41596-019-0201-3}.

\bibitem{Ciqtek}
\bibinfo{author}{Ciqtek}.
\newblock \bibinfo{title}{{Diamond quantum computer for education}}.
\newblock \bibinfo{howpublished}{\url{https://www.ciqtek.com/en/product/13},
  accessed 29/08/2021}.

\bibitem{SpinFlex}
\bibinfo{author}{Spin-Flex}.
\newblock \bibinfo{title}{{spinEDU -- Educational kit based on solid-state spin
  qubits in diamond for quantum technology}}.
\newblock \bibinfo{howpublished}{\url{https://spin-flex.com/spinedu/}, accessed
  29/08/2021}.

\bibitem{Childress2013}
\bibinfo{author}{Childress, L.} \& \bibinfo{author}{Hanson, R.}
\newblock \bibinfo{title}{{Diamond NV centers for quantum computing and quantum
  networks}}.
\newblock \emph{\bibinfo{journal}{MRS Bulletin}} \textbf{\bibinfo{volume}{38}},
  \bibinfo{pages}{134--138} (\bibinfo{year}{2013}).
\newblock \urlprefix\url{https://doi.org/10.1557/mrs.2013.20}.

\bibitem{Abobeih2018}
\bibinfo{author}{Abobeih, M.~H.} \emph{et~al.}
\newblock \bibinfo{title}{One-second coherence for a single electron spin
  coupled to a multi-qubit nuclear-spin environment}.
\newblock \emph{\bibinfo{journal}{Nature Communications}}
  \textbf{\bibinfo{volume}{9}}, \bibinfo{pages}{1--8} (\bibinfo{year}{2018}).
\newblock \urlprefix\url{https://doi.org/10.1038/s41467-018-04916-z}.

\bibitem{Abobeih2019}
\bibinfo{author}{Abobeih, M.} \emph{et~al.}
\newblock \bibinfo{title}{Atomic-scale imaging of a 27-nuclear-spin cluster
  using a quantum sensor}.
\newblock \emph{\bibinfo{journal}{Nature}} \textbf{\bibinfo{volume}{576}},
  \bibinfo{pages}{411--415} (\bibinfo{year}{2019}).
\newblock \urlprefix\url{https://doi.org/10.1038/s41586-019-1834-7}.

\bibitem{Bhaskar2020}
\bibinfo{author}{Bhaskar, M.~K.} \emph{et~al.}
\newblock \bibinfo{title}{Experimental demonstration of memory-enhanced quantum
  communication}.
\newblock \emph{\bibinfo{journal}{Nature}} \textbf{\bibinfo{volume}{580}},
  \bibinfo{pages}{60--64} (\bibinfo{year}{2020}).
\newblock \urlprefix\url{https://doi.org/10.1038/s41586-020-2103-5}.

\bibitem{Wan2020}
\bibinfo{author}{Wan, N.~H.} \emph{et~al.}
\newblock \bibinfo{title}{Large-scale integration of artificial atoms in hybrid
  photonic circuits}.
\newblock \emph{\bibinfo{journal}{Nature}} \textbf{\bibinfo{volume}{583}},
  \bibinfo{pages}{226--231} (\bibinfo{year}{2020}).
\newblock \urlprefix\url{https://doi.org/10.1038/s41586-020-2441-3}.

\bibitem{Bourassa2020}
\bibinfo{author}{Bourassa, A.} \emph{et~al.}
\newblock \bibinfo{title}{Entanglement and control of single nuclear spins in
  isotopically engineered silicon carbide}.
\newblock \emph{\bibinfo{journal}{Nature Materials}}
  \textbf{\bibinfo{volume}{19}}, \bibinfo{pages}{1319--1325}
  (\bibinfo{year}{2020}).
\newblock \urlprefix\url{https://doi.org/10.1038/s41563-020-00802-6}.

\bibitem{Freer2017}
\bibinfo{author}{Freer, S.} \emph{et~al.}
\newblock \bibinfo{title}{A single-atom quantum memory in silicon}.
\newblock \emph{\bibinfo{journal}{Quantum Science and Technology}}
  \textbf{\bibinfo{volume}{2}}, \bibinfo{pages}{015009} (\bibinfo{year}{2017}).
\newblock \urlprefix\url{https://doi.org/10.1088/2058-9565/aa63a4}.

\bibitem{Asaad2020}
\bibinfo{author}{Asaad, S.} \emph{et~al.}
\newblock \bibinfo{title}{Coherent electrical control of a single high-spin
  nucleus in silicon}.
\newblock \emph{\bibinfo{journal}{Nature}} \textbf{\bibinfo{volume}{579}},
  \bibinfo{pages}{205--209} (\bibinfo{year}{2020}).
\newblock \urlprefix\url{https://doi.org/10.1038/s41586-020-2057-7}.

\bibitem{Madzik2020}
\bibinfo{author}{Madzik, M.~T.} \emph{et~al.}
\newblock \bibinfo{title}{Conditional quantum operation of two exchange-coupled
  single-donor spin qubits in a mos-compatible silicon device}.
\newblock \emph{\bibinfo{journal}{Nature Communications}}
  \textbf{\bibinfo{volume}{12}}, \bibinfo{pages}{181} (\bibinfo{year}{2021}).
\newblock \urlprefix\url{https://doi.org/10.1038/s41467-020-20424-5}.

\bibitem{Hensen2020}
\bibinfo{author}{Hensen, B.} \emph{et~al.}
\newblock \bibinfo{title}{A silicon quantum-dot-coupled nuclear spin qubit}.
\newblock \emph{\bibinfo{journal}{Nature Nanotechnology}}
  \textbf{\bibinfo{volume}{15}}, \bibinfo{pages}{13--17}
  (\bibinfo{year}{2020}).
\newblock \urlprefix\url{https://doi.org/10.1038/s41565-019-0587-7}.

\bibitem{PhysRevB.79.075203}
\bibinfo{author}{Felton, S.} \emph{et~al.}
\newblock \bibinfo{title}{Hyperfine interaction in the ground state of the
  negatively charged nitrogen vacancy center in diamond}.
\newblock \emph{\bibinfo{journal}{Phys. Rev. B}} \textbf{\bibinfo{volume}{79}},
  \bibinfo{pages}{075203} (\bibinfo{year}{2009}).
\newblock \urlprefix\url{https://link.aps.org/doi/10.1103/PhysRevB.79.075203}.

\bibitem{phdthesis_Yossi}
\bibinfo{author}{Rosenzweig, Y.}
\newblock \emph{\bibinfo{title}{The Physics of Nitrogen-Vacancy Color Centers
  in Diamonds and their Interaction with External Fields}}.
\newblock Ph.D. thesis, \bibinfo{school}{Ben-Gurion University of the Negev}
  (\bibinfo{year}{2016}).
\newblock
  \urlprefix\url{https://www.bgu.ac.il/atomchip/Theses/Yossi_Rosenzweig_Msc_2016.pdf}.

\bibitem{phdthesis_claudia}
\bibinfo{author}{Avalos, C.~E.}
\newblock \emph{\bibinfo{title}{Detection and Polarization of Nuclear and
  Electron Spins using Nitrogen-Vacancy Centers}}.
\newblock Ph.D. thesis, \bibinfo{school}{University of California, Berkeley}
  (\bibinfo{year}{2014}).
\newblock
  \urlprefix\url{https://search.proquest.com/docview/1779999477?accountid=12763}.

\bibitem{Cohen1977}
\bibinfo{author}{Cohen-Tannoudji, C.}, \bibinfo{author}{Diu, B.} \&
  \bibinfo{author}{Lalo{\"e}, F.}
\newblock \emph{\bibinfo{title}{Quantum Mechanics, Volume 1: Basic Concepts,
  Tools, and Applications}} (\bibinfo{publisher}{John Wiley \& Sons},
  \bibinfo{year}{2019}).
\newblock
  \urlprefix\url{{https://www.wiley.com/en-us/Quantum+Mechanics\%2C+Volume+1\%3A+Basic+Concepts\%2C+Tools\%2C+and+Applications\%2C+2nd+Edition-p-9783527822713}}.

\bibitem{Note1}
\bibinfo{note}{$\Delta $ and $\Omega _{\protect \rm R}$ are defined in units of
  Hz.}

\bibitem{Vallabhapurapu2021}
\bibinfo{author}{Vallabhapurapu, H.~H.} \emph{et~al.}
\newblock \bibinfo{title}{{Fast coherent control of an NV-spin ensemble using a
  KTaO3 dielectric resonator at cryogenic temperatures}}.
\newblock \emph{\bibinfo{journal}{arXiv preprint arXiv:2105.06781}}
  (\bibinfo{year}{2021}).
\newblock \urlprefix\url{https://arxiv.org/abs/2105.06781}.

\bibitem{Note2}
\bibinfo{note}{The cancellation of the $\protect \genfrac {}{}{}1{1}{2}$-term
  derives from the ${\protect \rm sin}^2$-term. This is because a ${\protect
  \rm sin}^2$-function oscillates at twice the frequency as the corresponding
  ${\protect \rm sin}$-function.}

\bibitem{cvd}
\bibinfo{author}{Elementsix}.
\newblock \bibinfo{title}{{SC Plate CVD 2.6x2.6mm, 0.25mm thick,
  $\langle100\rangle$, P2}}.
\newblock
  \bibinfo{howpublished}{\url{https://e6cvd.com/us/application/general/sc-plate-cvd-2-6x2-6x0-25mm-100-p2.html},
  accessed 12/04/2019}.

\bibitem{DNV}
\bibinfo{author}{Elementsix}.
\newblock \bibinfo{title}{{DNV-B14 3.0x3.0mm, 0.5mm thick,
  $\langle100\rangle$}}.
\newblock
  \bibinfo{howpublished}{\url{https://e6cvd.com/uk/application/all/dnv-b14-\%203-0mmx3-0mm-0-5mm.html},
  accessed 08/01/2022}.

\bibitem{Acosta2010}
\bibinfo{author}{Acosta, V.~M.} \emph{et~al.}
\newblock \bibinfo{title}{Temperature dependence of the nitrogen-vacancy
  magnetic resonance in diamond}.
\newblock \emph{\bibinfo{journal}{Phys. Rev. Lett.}}
  \textbf{\bibinfo{volume}{104}}, \bibinfo{pages}{070801}
  (\bibinfo{year}{2010}).
\newblock \urlprefix\url{https://doi.org/10.1103/PhysRevLett.104.070801}.

\bibitem{van1997dependences}
\bibinfo{author}{Van~Wyk, J.}, \bibinfo{author}{Reynhardt, E.},
  \bibinfo{author}{High, G.} \& \bibinfo{author}{Kiflawi, I.}
\newblock \bibinfo{title}{The dependences of esr line widths and spin-spin
  relaxation times of single nitrogen defects on the concentration of nitrogen
  defects in diamond}.
\newblock \emph{\bibinfo{journal}{Journal of Physics D: Applied Physics}}
  \textbf{\bibinfo{volume}{30}}, \bibinfo{pages}{1790} (\bibinfo{year}{1997}).
\newblock \urlprefix\url{https://doi.org/10.1088/0022-3727/30/12/016}.

\bibitem{Smeltzer_2011}
\bibinfo{author}{Smeltzer, B.}, \bibinfo{author}{Childress, L.} \&
  \bibinfo{author}{Gali, A.}
\newblock \bibinfo{title}{{13C hyperfine interactions in the nitrogen-vacancy
  centre in diamond}}.
\newblock \emph{\bibinfo{journal}{New Journal of Physics}}
  \textbf{\bibinfo{volume}{13}}, \bibinfo{pages}{025021}
  (\bibinfo{year}{2011}).
\newblock \urlprefix\url{https://doi.org/10.1088/1367-2630/13/2/025021}.

\bibitem{Note3}
\bibinfo{note}{As this results in a lower photoluminescence signal, we use
  continuous ODMR to compensate for the decrease of signal strength, where the
  laser is on throughout the whole lock-in cycle, while the MW is pulsed on
  only during the first half.}

\bibitem{Pham2013MagneticFS}
\bibinfo{author}{Pham, L.~M.}
\newblock \emph{\bibinfo{title}{Magnetic Field Sensing with Nitrogen-Vacancy
  Color Centers in Diamond}}.
\newblock Ph.D. thesis, \bibinfo{school}{Harvard University}
  (\bibinfo{year}{2013}).
\newblock \urlprefix\url{http://nrs.harvard.edu/urn-3:HUL.InstRepos:11051173}.

\bibitem{Barry2020}
\bibinfo{author}{Barry, J.~F.} \emph{et~al.}
\newblock \bibinfo{title}{{Sensitivity optimization for NV-diamond
  magnetometry}}.
\newblock \emph{\bibinfo{journal}{Reviews of Modern Physics}}
  \textbf{\bibinfo{volume}{92}}, \bibinfo{pages}{015004}
  (\bibinfo{year}{2020}).
\newblock \urlprefix\url{https://doi.org/10.1103/RevModPhys.92.015004}.

\bibitem{Grezes2016}
\bibinfo{author}{Gr{\`e}zes, C.}
\newblock \emph{\bibinfo{title}{Experiment 1 (Write): Coherent Storage of Qubit
  States into a Spin Ensemble}}, \bibinfo{pages}{93--132}
  (\bibinfo{publisher}{Springer International Publishing},
  \bibinfo{address}{Cham}, \bibinfo{year}{2016}).
\newblock \urlprefix\url{https://doi.org/10.1007/978-3-319-21572-3_4}.

\bibitem{Nielsen2002}
\bibinfo{author}{Nielsen, M.~A.} \& \bibinfo{author}{Chuang, I.}
\newblock \emph{\bibinfo{title}{Quantum Computation and Quantum Information:
  10th Anniversary Edition}} (\bibinfo{publisher}{Cambridge University Press},
  \bibinfo{year}{2010}).
\newblock \urlprefix\url{https://doi.org/10.1017/CBO9780511976667}.

\bibitem{Hahn1950}
\bibinfo{author}{Hahn, E.~L.}
\newblock \bibinfo{title}{Spin echoes}.
\newblock \emph{\bibinfo{journal}{Physical Review}}
  \textbf{\bibinfo{volume}{80}}, \bibinfo{pages}{580--594}
  (\bibinfo{year}{1950}).
\newblock \urlprefix\url{https://doi.org/10.1103/PhysRev.80.580}.

\bibitem{Childress281}
\bibinfo{author}{Childress, L.} \emph{et~al.}
\newblock \bibinfo{title}{Coherent dynamics of coupled electron and nuclear
  spin qubits in diamond}.
\newblock \emph{\bibinfo{journal}{Science}} \textbf{\bibinfo{volume}{314}},
  \bibinfo{pages}{281--285} (\bibinfo{year}{2006}).
\newblock \urlprefix\url{https://doi.org/10.1126/science.1131871}.

\bibitem{PhysRevB.82.201201}
\bibinfo{author}{Stanwix, P.~L.} \emph{et~al.}
\newblock \bibinfo{title}{Coherence of nitrogen-vacancy electronic spin
  ensembles in diamond}.
\newblock \emph{\bibinfo{journal}{Phys. Rev. B}} \textbf{\bibinfo{volume}{82}},
  \bibinfo{pages}{201201} (\bibinfo{year}{2010}).
\newblock \urlprefix\url{https://doi.org/10.1103/PhysRevB.82.201201}.

\end{thebibliography}

\clearpage
\onecolumngrid
\appendix
\section*{Appendix A: Design of the electromagnet}\label{AppA}

\begin{figure*}[hbt]
	\centering 
        \includegraphics[width=\textwidth]{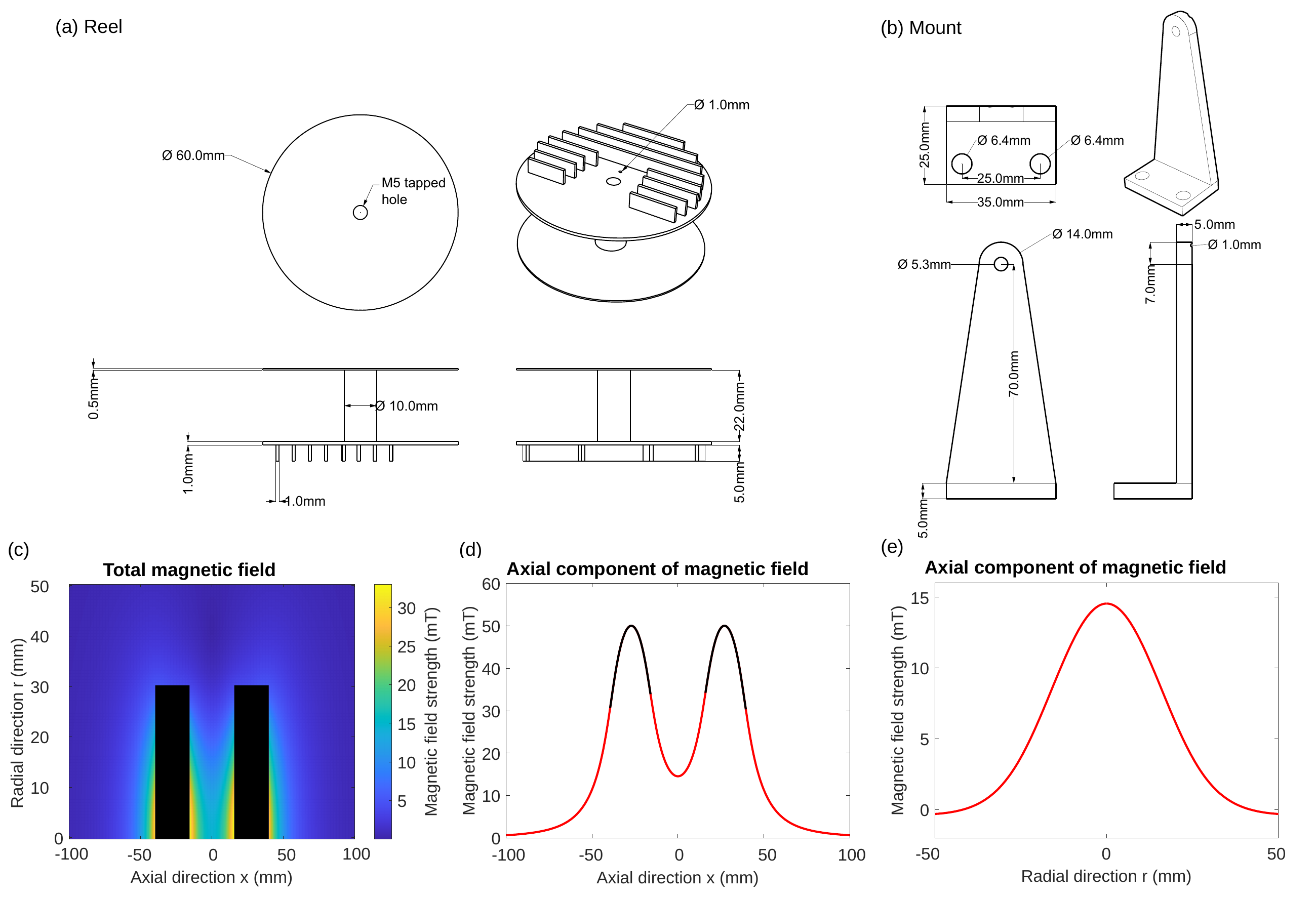}
        \caption{Design details of the electromagnet. 
        (a) Sketch of the reel. The M5 tapped hole is used to fasten the reel to the mount [see panel (b)] using a stainless steel screw. In addition we have added 12~mm long supermendur cylinders with 4~mm diameter into the holes to slightly increase the magnetic field (optional). The 1~mm diameter hole (drilled at a 45$^\circ$ angle) allows the inner end of the copper wire to pass through. The copper wire we used is American Wire Gauge (AWG) No.22 with 0.644~mm diameter and 52.96~m$\Omega$ resistance per meter. Several 1~mm thick and 5~mm long cooling fences allow for better heat dissipation. We chose brass as material due to its non-magnetic properties and ease of manufacturability. 
        (b) Sketch of the L-shaped mount. The 5.3~mm diameter hole is used to mount the reel, while the 1~mm diameter trench allows the copper wire from the reel to pass through. The connected reel and mount are screwed onto the optical breadboard via two 6.4~mm diameter holes on the L-shaped mount. 
        (c-e) Magnetic field simulation results. The results were obtained by calculating the magnetic field produced by the wire loops using the Biot-Savart law in matlab. The simulation is done under the condition that the distance between the two reels is 32~mm and a 1~A current is passing through the coils. The supermendur cylinders were not included in the simulations. Two black areas in (c) and (d) indicate the physical locations of the coils. (c) shows a 2D map of the total magnetic field along the radial and axial directions of the coils, with (d) showing the axial magnetic field component along the axial direction at $r=0$~m and (e) along the radial direction at $x=0$~m. The magnetic field at the center is 14.5~mT [see also dip in (d) and peak in (e)] and the simulated power consumption is 10.6~W. The real power consumption is slightly higher due to the heat generated by the coils causing an increase in the copper wire resistance. The negative values in (e) indicate that the axial component of the magnetic field is pointing along the negative x direction. The gradient of the magnetic field is smallest at the center between the two reels, providing the most homogeneous magnetic field in both strength and direction.}
        \label{electromagnets}
\end{figure*}

\end{document}